\documentclass[12pt,a4paper]{article}
\usepackage{amssymb}

\usepackage[left=5em]{geometry}
\usepackage{amsmath}
\usepackage{graphicx}

\title{Inconsistences in Interacting Agegraphic Dark Energy Models}
\author{\textbf{Cheng-Yi Sun\footnote{cysun@mailis.gucas.ac.cn; ddscy@163.com}\
$^{,a}$} and \textbf{Yu Song$^{a}$
}\\ \\
 {$^a$\small Institute of Modern Physics, Northwest University,}\\
     \small Xian 710069, P.R. China.}

\begin{document}
\maketitle
\begin{abstract}
It is found that the origin agegraphic dark energy tracks the matter
in the matter-dominated epoch and then the subsequent
dark-energy-dominated epoch becomes impossible. It is argued that
the difficulty can be removed when the interaction between the
agegraphic dark energy and dark matter is considered. In the note,
by discussing three different interacting models, we find that the
difficulty still stands even in the interacting models. Furthermore,
we find that in the interacting models, there exists the other
serious inconsistence that the existence of the
radiation/matter-dominated epoch contradicts the ability of
agegraphic dark energy in driving the accelerated expansion. The
contradiction can be avoided in one of the three models if some
constraints on the parameters hold.
\end{abstract}



\section{Introduction}

Increasing evidence suggests that the expansion of our universe is
being accelerated \cite{Supernova,WMAP,LSS}. Within the framework of
the general relativity, the acceleration can be phenomenally
attributed to the existence of a mysterious exotic component with
negative pressure, namely the dark energy \cite{dark energy1,dark
energy2}. However, we know little about the nature of dark energy.
The most nature, simple and important candidate for dark energy is
the Einstein's cosmological constant, which can fit the observations
well so far. But the cosmological constant is plagued with the
well-known fine-tuning and cosmic coincidence problems \cite{dark
energy1,dark energy2}. Dark energy has become one of the most active
fields in the modern cosmology.

Recently, the so-called agegraphic dark energy model is suggested
\cite{0707.4049}. The energy density of agegraphic dark energy is
given by \cite{0707.4049}
\begin{equation}
  \label{ADE}
  \rho_D=\frac{3n^2M_p^2}{T^2}.
\end{equation}
Here $M_p=(8\pi G)^{-1/2}$ and $T$ is chosen to be the age of our
universe
\begin{equation}
  \label{T}
  T=\int^t_0{dt'}=\int^a_0{\frac{da}{Ha}},
\end{equation}
where $a$ is the scale factor of our universe, $H\equiv\dot{a}/a$ is
the Hubble parameter and a dot denotes the derivative with respect
to cosmic time. However, it is found that the agegraphic dark energy
proposed in \cite{0707.4049} tracks the matter in the
matter-dominated epoch \cite{0708.0884}. This can be understood
easily \cite{1005.2466}. In the matter-dominated epoch, $a\propto
t^{2/3}$. Then we have $\rho_D\propto a^{-3}$. Since the energy
density of matter $\rho_m\propto a^{-3}$, $\rho_D$ tracks $\rho_m$
in the matter-dominated epoch and the dark-energy-dominated epoch
becomes impossible. This is of course unacceptable.

Two ways out of the difficulty are suggested. The first one is to
replace $T$ with $T+\delta$ \cite{0707.4049}, where $\delta$ is a
constant with dimension of time. The second one is the so-called new
agegraphic dark energy by replacing $T$ with the conformal time
$\eta$ \cite{0708.0884}. Both of the ways change Eq.(\ref{ADE}). In
\cite{0707.4052}, it is argued that the difficulty in the origin
version can also be removed when the interaction between the
agegraphic dark energy and matter is considered.

In note, the interacting agegraphic dark energy models with three
different forms of interaction are considered respectively. We find
that in the model with interaction proportional to the energy
density of matter, $\rho_D$ still tracks $\rho_m$ in the
matter-dominated epoch. Furthermore, we find that the existence of
the matter-dominated epoch contradicts the ability of agegraphic
dark energy in driving the accelerated expansion even in the
interacting models.

The paper is organized as follows. In the next section, we will
discuss the difficulties of the agegraphic dark energy. In
Sec.\ref{SecIADE}, we will recall the analysis in \cite{0707.4052}
which tells us that the agegraphic dark energy will not track the
matter during the matter-dominated epoch if the interaction between
agegraphic dark energy and dark matter is considered. In
Sec.\ref{SecTBIADE}, we will show that there exists the other
inconsistence in the interacting models. Finally, Conclusions and
Discussions are given.

\section{Inconsistences in Agegraphic Dark Energy Model}
\label{SecTBADE}

Considering the flat Friedmann-Robertson-Walk universe with the
agegraphic dark energy and pressureless matter, the corresponding
Friedmann equation is
\begin{equation}
  \label{FE}
  H^2=\frac{1}{3M_p^2}(\rho_m+\rho_D).
\end{equation}
By defining
\begin{equation}
  \label{Omega}
  \rho_c=3H^2M_p^2,\quad\Omega_m=\frac{\rho_m}{\rho_c},\quad\Omega_D=\frac{\rho_D}{\rho_c},
\end{equation}
we may rewrite the Friedmann equation as
\begin{equation}
  \label{rFE}
  1=\Omega_m+\Omega_D.
\end{equation}
And from Eq.(\ref{ADE}), we can easily find that
\begin{equation}
  \label{OmegaD}
  \Omega_D=\frac{n^2}{H^2T^2}.
\end{equation}
The conservation laws of the agegraphic dark energy and matter are
respectively
\begin{align}
  \label{CLm}
  \dot{\rho}_m+3H\rho_m&=0,\\
  \label{CLD}
  \dot{\rho}_D+3H(1+w_D)\rho_D&=0,
\end{align}
where $w_D$ is the equation-of-state (EoS) parameter of the
agegraphic dark energy. Taking the derivative of Eq.(\ref{ADE}) with
respect to the cosmic time $t$, we can get
\begin{equation}
  \label{dRhoDdt}
  \dot{\rho}_D+H\frac{2\sqrt{\Omega_D}}{n}\rho_D=0.
\end{equation}
Comparing the equation with Eq(\ref{CLD}), we get \cite{0707.4049}
\begin{equation}
  \label{wD}
  w_D=-1+\frac{2}{3n}\sqrt{\Omega_D}.
\end{equation}
From Eq.(\ref{OmegaD}) and using Eqs.(\ref{FE}), (\ref{Omega}),
(\ref{CLm}) and (\ref{dRhoDdt}), we can have \cite{0707.4049}
\begin{equation}
  \label{dOmegaDdlna}
  \Omega'_D=\Omega_D(1-\Omega_D)\Big(3-\frac{2}{n}\sqrt{\Omega_D}\Big),
\end{equation}
where a prime denotes the derivative with respect to the e-folding
time $N=\ln a$. The evolution of $\Omega_D$ governed by
Eq.(\ref{dOmegaDdlna}) has been analyzed in Ref.\cite{0707.4049} and
it is found that the agegraphic dark energy model works well
\cite{0707.4049}.

However, it is found in \cite{0707.4049,0708.0884} that there exists
an implicit inconsistence in the agegraphic dark energy model. In
the matter-dominated epoch with $\Omega_D\ll 1$, the Friedmann
equation approximately becomes
\begin{equation}
  \label{appFE}
  H^2\simeq\frac{1}{3M_p^2}\rho_m.
\end{equation}
Together with Eq.(\ref{CLm}), we have $a\propto t^{2/3}$. Then we
have
\[
  H=\frac{2}{3t}.
\]
Substituting the result into Eq.(\ref{OmegaD}), we have
\begin{equation}
  \label{OmegaDC}
  \Omega_{DI}=\frac{9n^2}{4}.
\end{equation}
Hereafter we use the subscript $I$ to denote the initial value of
$\Omega_D$ when deep in the matter-dominated epoch. It can be easily
checked that Eq.(\ref{OmegaDC}) is also a critical point of
Eq.(\ref{dOmegaDdlna}). In fact, there exist the other two critical
points of Eq.(\ref{dOmegaDdlna}),
\begin{align}
  \label{CP0}
  \Omega_{D1}=0,\\
  \label{CP1}
  \Omega_{D2}=1.
\end{align}
Obviously, Eq.(\ref{CP0}) is unstable. Since Eq.(\ref{OmegaDC}) is
the value of $\Omega_D$ deep in the matter-dominated epoch, then we
have
\begin{equation}
  \label{OmegaDCll1}
  \Omega_{DI}=\frac{9n^2}{4}\ll1.
\end{equation}
With this result, it can be checked easily that Eq.(\ref{OmegaDC})
is an attractor, while Eq.(\ref{CP1}) is unstable. This implies
$\Omega_D\rightarrow\frac{9n^2}{4}\ll1$ as $\ln a\rightarrow+\infty$
and the subsequent dark-energy-dominated epoch becomes impossible.

Furthermore, we find that Eq.(\ref{OmegaDC}) indicates the other
serious problem. This can be easily shown as follows. From
Eq.(\ref{OmegaDCll1}), we have
\[
  n\ll 1.
\]
On the other hand, Eq.(\ref{wD}) tells us that the necessary
condition for agegraphic dark energy to drive the accelerated
expansion is
\begin{equation}
  \label{ng1}
  n>1.
\end{equation}
So the matter-dominated epoch contradicts the ability of agegraphic
dark energy in driving the accelerated expansion.

A similar conclusion is also given in Ref.\cite{0708.2910}. By
extending the discussion of agegraphic dark energy model to include
the radiation-dominated epoch, the author in \cite{0708.2910} noted
that the bound imposed on the fractional dark energy density
parameter $\Omega_D<0.1$ during the big bang nucleosynthesis (BBN)
requires
\begin{equation}
  \label{nBBN}
  n<\frac{1}{6}.
\end{equation}
The contradiction between Eqs(\ref{ng1}) and (\ref{nBBN}) is
obvious. Then it is interesting for us to explore whether the
contradiction can be solved when the interaction is involved.

\section{Interacting Agegraphic Dark Energy Models}
\label{SecIADE}

In \cite{0707.4052}, it is shown that the agegraphic dark energy
will not track the matter when the interaction between the
agegraphic dark energy and matter is considered. In the section, we
will recall the analysis in \cite{0707.4052}. In the next section,
we will show that, actually, there still exist some inconsistences
even in the interacting models.

The conservation laws of the agegraphic dark energy and matter are
respectively \cite{0707.4052}
\begin{align}
  \label{CLIm}
  \dot{\rho}_m+3H\rho_m&=Q,\\
  \label{CLID}
  \dot{\rho}_D+3H(1+w_D)\rho_D&=-Q,
\end{align}
where $Q$ denotes the phenomenological interaction term. Comparing
Eq.(\ref{dRhoDdt}) and Eq.(\ref{CLID}), we have \cite{0707.4052}
\begin{equation}
  \label{wDI}
  w_D=-1+\frac{2}{3n}\sqrt{\Omega_D}-\frac{Q}{3H\rho_D}.
\end{equation}
If $Q=0$, the equation reduces to Eq.(\ref{wD}). From
Eqs.(\ref{FE}), (\ref{dRhoDdt}) and (\ref{CLIm}), we have
\begin{equation}
  \label{dadtt}
  \frac{\ddot{a}}{a}=-\frac{4\pi
  G}{3}\Big(\rho_m-2\rho_D+\frac{2\sqrt{\Omega_D}}{n}\rho_D-\frac{Q}{H}\Big).
\end{equation}

From Eq.(\ref{OmegaD}) and using Eqs.(\ref{FE}), (\ref{Omega}),
 (\ref{dRhoDdt}) and (\ref{CLIm}), we can obtain the evolving
equation of $\Omega_D$ with interaction \cite{0707.4049}
\begin{equation}
  \label{dOmegaDdlnaQ}
  \Omega'_D=\Omega_D\Big[(1-\Omega_D)\Big(3-\frac{2}{n}\sqrt{\Omega_D}\Big)-\frac{Q}{3M_p^2H^3}\Big].
\end{equation}
If $Q=0$, this equation reduces to Eq.(\ref{dOmegaDdlna}).

In \cite{0707.4052}, Eq.(\ref{dOmegaDdlnaQ}) has been solved
numerically with the initial condition $\Omega_{D0}=0.7$, and the
reasonable evolution of $\Omega_D$ has been shown that the
agegraphic dark energy is negligible in the past and eventually
dominates the evolution of our universe. Then it seems to be
reasonable to conclude that the inconsistence in the non-interacting
agegraphic dark energy has been removed in the interacting models.
However, as shown below, we find that there still exist the
inconsistences in the interacting agegraphic dark energy models.

\section{Inconsistences in Interacting Agegraphic Dark Energy Models}
\label{SecTBIADE}

In the section, owing to the lack of the knowledge of micro-origin
of the interaction, we simply consider three forms of the
interaction
\begin{equation}
  \label{Q}
  Q=3\beta H\rho_m,\ 3\alpha H\rho_D,\ 3\gamma H(\rho_m+\rho_D),
\end{equation}
which are used in the literature most often
\cite{0707.4052,InteractionQ1,InteractionQ2,InteractionQ3,Wei,ZKGuo}.
Here $\beta,\alpha$ and $\gamma$ are positive constants.

\subsection{$Q=3\beta H\rho_m$}
\label{SubSecBeta}

Firstly, we consider the interacting agegraphic dark energy model
with $Q=3\beta H\rho_m$. Then Eq.(\ref{dOmegaDdlnaQ}) reads
\begin{equation}
  \label{dOmegaDdlnaQ2}
  \Omega'_D=\Omega_D(1-\Omega_D)\Big[3(1-\beta)-\frac{2}{n}\sqrt{\Omega_D}\Big)\Big].
\end{equation}
And the conservation law of matter reads
\begin{equation}
  \label{CLIm2}
  \dot{\rho}_m+3H\rho_m=3H\beta\rho_m.
\end{equation}
From the above equation, we can easily obtain
\begin{equation}
  \label{rhoma2}
  \rho_m\propto a^{-3(1-\beta)}.
\end{equation}
Then, in the matter-dominated epoch with $\Omega_D\ll 1$, from
Eq.(\ref{appFE}), we have
\begin{equation}
  \label{HIADE2}
  a\propto t^{\frac{2}{3(1-\beta)}}\Rightarrow H=\frac{2}{3(1-\beta)}\frac{1}{t}.
\end{equation}
Substituting the equation into Eq.(\ref{OmegaD}), we have
\begin{equation}
  \label{OmegaDC2}
  \Omega_{DI}=\Big[\frac{3(1-\beta)n}{2}\Big]^2.
\end{equation}
This is the initial value of $\Omega_D$ in the model with $Q=3\beta
H\rho_m$ when deep in the matter-dominated epoch. It can be checked
easily that Eq.(\ref{OmegaDC2}) is also a critical point of
Eq.(\ref{dOmegaDdlnaQ2}). And the other two critical points of
Eq.(\ref{dOmegaDdlnaQ2}) are given in Eqs.(\ref{CP0}) and
(\ref{CP1}) respectively. Since deep in the matter-dominated epoch
\begin{equation}
  \label{OmegaDC2ll1}
  \Omega_{DI}\ll1,
\end{equation}
then we can find that Eq.(\ref{OmegaDC2}) is an attractor of
Eq.(\ref{dOmegaDdlnaQ2}) while the other two critical points of
Eq.(\ref{dOmegaDdlnaQ2}) are unstable. This implies
$\Omega_D\rightarrow\Big[\frac{3(1-\beta)n}{2}\Big]^2\ll1$ as $\ln
a\rightarrow+\infty$ and the subsequent dark-energy-dominated epoch
becomes impossible.

Furthermore, we find that the contradiction between the
matter-dominated epoch and the ability of agegraphic dark energy in
driving the accelerated expansion also exists in the interacting
model. Let us show it. From Eq.(\ref{dadtt}), in the
matter-dominated epoch, approximately we have
\[
  \frac{\ddot{a}}{{a}}\simeq-\frac{4\pi G}{3}(1-3\beta)\rho_m,
\]
since $\rho_D\ll\rho_m$. Then we must have
\begin{equation}
  \label{beta}
  \beta<\frac{1}{3},
\end{equation}
since, if $\beta>\frac{1}{3}$, the expansion of the universe during
the matter-dominated epoch would be accelerated and then the
observed large scale structure of the universe could not be formed.
Substituting $\beta<\frac{1}{3}$ into Eq.(\ref{OmegaDC2}), we have
\begin{equation}
  \label{OmegaDLessN}
  \Omega_{DI}>n^2.
\end{equation}
Then together with Eq.(\ref{OmegaDC2ll1}), we have
\begin{equation}
  \label{nQ2}
  n\ll1,
\end{equation}
On the other hand, in the dark-energy-dominated epoch, since
$\Omega_D\simeq1$ and $\rho_m\ll\rho_D$, from Eq.(\ref{dadtt}) we
have
\begin{equation}
  \label{dadttDDQ2}
  \frac{\ddot{a}}{a}\simeq-\frac{8\pi
  G}{3}\Big(\frac{1}{n}-1\Big)\rho_D,
\end{equation}
where $Q=3\beta H\rho_m$ has been used. Then in order for the
agegraphic dark energy to drive the accelerated expansion, we must
have
\[
  n>1.
\]
Obviously, this result contradicts Eq.(\ref{nQ2}). So the
contradiction between the matter-dominated epoch and the ability of
agegraphic dark energy in driving the accelerated expansion still
stands in the interacting model with $Q=3\beta H\rho_m$.

\subsection{$Q=3\alpha H\rho_D$}

Secondly, we consider the case of $Q=3\alpha H\rho_D$. Then
Eq.(\ref{dOmegaDdlnaQ}) reads
\begin{equation}
  \label{dOmegaDdlnaQ1}
  \Omega'_D=\Omega_D\Big[(1-\Omega_D)\Big(3-\frac{2}{n}\sqrt{\Omega_D}\Big)-3\alpha\Omega_D\Big].
\end{equation}
In the matter-dominated epoch, Eq.(\ref{CLIm}) reads approximately
\begin{equation}
  \label{CLIm1}
  \dot{\rho}_m+3H\rho_m\simeq0,
\end{equation}
since $\rho_D\ll\rho_m$. Then in the matter-dominated epoch,
approximately we have
\begin{equation}
  \label{rhoma1}
  \rho_m\propto a^{-3},
\end{equation}
So from Eqs.(\ref{appFE}) and (\ref{rhoma1}), we  have
\begin{equation}
  \label{HIADE1}
  a\propto t^{\frac{2}{3}}\Rightarrow H=\frac{2}{3t}.
\end{equation}
Substituting the equation into Eq.(\ref{OmegaD}), we have
\begin{equation}
  \label{OmegaDC1}
  \Omega_{DI}=\frac{9n^2}{4}.
\end{equation}
Then, when deep in the matter-dominated epoch, we have the same
initial value of $\Omega_D$ as in the non-interacting model. But the
case is different. Here, due to the interaction, Eq.(\ref{OmegaDC1})
is not the critical point of the evolving equation
(\ref{dOmegaDdlnaQ1}). Then it seems that the tracking behavior of
agegraphic dark energy during the matter-dominated epoch is
eliminated, and eventually the agegraphic dark energy will become
dominated. However, we find this problem still stands. As in the
non-interacting model, here we also have
\begin{equation}
  \label{OmegaDC1ll1}
  \Omega_{DI}=\frac{9n^2}{4}\ll1.
\end{equation}
Then, using this result, from Eq.(\ref{dOmegaDdlnaQ1}), we find
\begin{equation}
  \label{dOmegaDdlnaQ1l0}
  \Omega'_D<0, \quad\text{for}\ \Omega_{DI}\le\Omega_D\le1.
\end{equation}
Here we have used $\alpha>0$. Eq.(\ref{dOmegaDdlnaQ1l0}) tells us
that $\Omega_D$ will never become larger than $\frac{9n^2}{4}$ and
will approach a value less than $\Omega_{DI}=\frac{9n^2}{4}$ as $\ln
a\rightarrow+\infty$, and consequently the subsequent
dark-energy-dominated epoch is impossible.

Now let us show whether the contradiction between the
matter-dominated epoch and the ability of agegraphic dark energy in
driving the accelerated expansion exists in the case.
Eq.(\ref{OmegaDC1ll1}) implies
\begin{equation}
  \label{nll1Q1}
  n\ll1.
\end{equation}
On the other hand, in the dare-energy-dominated epoch, since
$\Omega_D\simeq1$ and $\rho_D\gg\rho_m$, from Eq.(\ref{dadtt}),
approximately we have
\[
  \frac{\ddot{a}}{a}\simeq-\frac{4\pi
  G}{3}\Big(-2+\frac{2}{n}-3\alpha\Big)\rho_D.
\]
Then, in order for the agegraphic dark energy to drive the
accelerated expansion, we must have
\begin{equation}
  \label{nQ1alpha}
  \frac{1}{n}<1+\frac{3}{2}\alpha.
\end{equation}
Together with Eq.(\ref{nQ1alpha}), we have
\begin{equation}
  \label{alpha}
  \frac{2}{2+3\alpha}<n\ll1.
\end{equation}
Then in the interacting model, we can remove the contradiction
between the matter-dominated epoch and the ability of agegraphic
dark energy in driving the accelerated expansion if Eq.(\ref{alpha})
holds.

\subsection{$Q=3\gamma H(\rho_D+\rho_m)$}

Finally, we consider the case of $Q=3\gamma H(\rho_D+\rho_m)$. In
the case Eq.(\ref{dOmegaDdlnaQ}) reads
\begin{equation}
  \label{dOmegaDdlnaQ3}
  \Omega'_D=\Omega_D\Big[(1-\Omega_D)\Big(3-\frac{2}{n}\sqrt{\Omega_D}\Big)-3\gamma\Big].
\end{equation}In
the matter-dominated epoch, since $\rho_D\ll\rho_m$, the
conservation law of matter reads approximately
\begin{equation}
  \label{CLIm3}
  \dot{\rho}_m+3H(1-\gamma)\rho_m\simeq0.
\end{equation}
Then we have
\begin{equation}
  \label{rhoma2}
  \rho_m\propto a^{-3(1-\gamma)}.
\end{equation}
Together with Eq.(\ref{appFE}), in the matter-dominated epoch we
have
\begin{equation}
  \label{HIADE2}
  a\propto t^{\frac{2}{3(1-\gamma)}}\Rightarrow H=\frac{2}{3(1-\gamma)}\frac{1}{t}.
\end{equation}
Substituting the equation into Eq.(\ref{OmegaD}), we have
\begin{equation}
  \label{OmegaDC3}
  \Omega_{DI}=\Big[\frac{3(1-\gamma)n}{2}\Big]^2.
\end{equation}
Due to the interaction, Eq.(\ref{OmegaDC3}) is not the critical
point of Eq.(\ref{dOmegaDdlnaQ3}). However, since in the
matter-dominated epoch
\begin{equation}
  \label{OmegaDC3ll1}
  \Omega_{DI}=\Big[\frac{3(1-\gamma)n}{2}\Big]^2\ll1,
\end{equation}
then from Eq.(\ref{dOmegaDdlnaQ3}) we find
\begin{equation}
  \label{dOmegaDdlnaQ3l0}
  \Omega'_D<0, \quad \text{for}\ \Omega_{DI}\le\Omega_D\le1.
\end{equation}
Here $\gamma>0$ has been used. As in the case of $Q=3\alpha
H\rho_D$, Eq.(\ref{dOmegaDdlnaQ3l0}) tells us that $\Omega_D$ will
approach a value smaller than
$\Omega_{DI}=\Big[\frac{3(1-\gamma)n}{2}\Big]^2$ as $\ln
a\rightarrow+\infty$ and the subsequent dark-energy-dominated epoch
becomes impossible.

Furthermore, we find that as in the case of $Q=3\beta H\rho_m$, in
the model with $Q=3\gamma H(\rho_D+\rho_m)$, the matter-dominated
epoch contradicts the ability of agegraphic dark energy in driving
the accelerated expansion, too. Let us show it. In the
matter-dominated epoch, since $\rho_D\ll\rho_m$, from
Eq.(\ref{dadtt}), approximately we have
\begin{equation}
  \label{dadttMDQ3}
  \frac{\ddot{a}}{a}\simeq-\frac{4\pi G}{3}(1-3\gamma)\rho_m,
\end{equation}
Then we must have
\begin{equation}
  \label{gamma}
  \gamma<\frac{1}{3},
\end{equation}
in order for the expansion of the universe to be decelerated to form
the large scale structure during the matter-dominated epoch. Using
Eqs.(\ref{gamma})  and (\ref{OmegaDC3ll1}) we have
\begin{equation}
  \label{nll1Q3}
  n\ll1.
\end{equation}
On the other hand, in the dark-energy-dominated epoch, since
$\rho_D\gg\rho_m$ and $\Omega_D\simeq1$, from Eq.(\ref{dadtt}),
approximately we have
\begin{equation}
  \label{dadttDDQ3}
  \frac{\ddot{a}}{a}\simeq-\frac{4\pi
  G}{3}\Big(-2+\frac{2}{n}-3\gamma\Big)\rho_D,
\end{equation}
where $Q=3\gamma H(\rho_D+\rho_m)\simeq3\gamma H\rho_D$ has been
used. Then the necessary condition for agegraphic dark energy to
drive the accelerated expansion is
\begin{equation}
  \label{gammaAndN}
  n>\frac{2}{2+3\gamma}.
\end{equation}
Together with Eq.(\ref{gamma}), we have
\begin{equation}
  \label{nQ3}
  n>\frac{2}{3}.
\end{equation}
Roughly, it seems that Eq.(\ref{nQ3}) may not contradict
Eq.(\ref{nll1Q3}). But the contradiction between Eq.(\ref{nQ3}) and
Eq.(\ref{nBBN}) is obvious. Here we note that the condition
(\ref{nBBN}) imposed by BBN on the agegraphic dark energy model is
not effected by the interaction between dark energy and matter,
since both agegraphic dark energy and matter are negligible during
the radiation-dominated epoch. So, in the interacting agegraphic
dark energy model with $Q=3\gamma H(\rho_D+\rho_m)$, the existence
of the radiation-dominated epoch contradicts the ability of
agegraphic dark energy in driving the accelerated expansion.

\section{Conclusions and Discussions}

In this note, we recall the inconsistence in the origin agegraphic
dark energy model that the agegraphic dark energy tracks the matter
in the matter-dominated epoch. And furthermore, we point out that
there is the other more serious inconsistence in the model that the
matter-dominated epoch contradicts the ability of agegraphic dark
energy in driving the accelerated expansion.

Then, by considering three kinds of phenomenological interaction
between the agegraphic dark energy and matter, we analyze the
interacting agegraphic dark energy models. We find that in the dark
energy model with interaction $Q=3\beta H\rho_m$, the agegraphic
dark energy still tracks the matter during the matter-dominated
epoch, and the contradiction between the existence of the
matter-dominated epoch and the ability of the agegraphic dark energy
in driving the accelerated expansion also exists. In the models with
$Q=3\alpha H\rho_D$ and $Q=3\alpha H\rho_D$, it is still impossible
for agegraphic dark energy to become dominated. And in the model
with $Q=3\alpha H\rho_D$ the contradiction between the existence of
the matter-dominated epoch and the ability of the agegraphic dark
energy in driving the accelerated expansion can be avoided if the
condition (\ref{alpha}) holds. But in the model with $Q=3\gamma
H(\rho_m+\rho_D)$, the ability of the agegraphic dark energy in
driving the accelerated expansion contradicts the bound imposed by
BBN on the agegraphic dark energy.

Then it seems that none of the three interacting agegraphic dark
energy models can be taken as serious candidate for realistic dark
energy. In Ref.\cite{1010.0567}, the authors studied the interacting
agegraphic dark energy model by using a general form of interaction
$Q=3H(\alpha\rho_D+\beta\rho_m)$. In the matter-dominated epoch with
$\rho_D\ll\rho_m$, the interaction reduces to
$Q\simeq3H\beta\rho_m$. So $\beta$ and $n$ should satisfy the
constraints (\ref{beta}) and (\ref{nQ2}) respectively, since the two
constraints are obtained by analyzing the model with
$Q=3H\beta\rho_m$ during the matter-dominated epoch. Similarly,
since the general interaction form reduces to
$Q\simeq3H\alpha\rho_D$ in the dark-energy-dominated epoch, $\alpha$
and $n$ should satisfy the condition (\ref{alpha}). However, it can
be checked easily that the values of parameters used in
Ref.\cite{1010.0567} does not satisfy Eq.(\ref{alpha}). Then there
exist implicit inconsistences in the interacting agegraphic dark
energy models analyzed in Ref.\cite{1010.0567}, although the authors
obtained the reasonable behaviors of $\Omega_D$ and $\Omega_m$ by
solving the evolving equations numerically. This would not be
confusing or astonishing if we recall that even in the
non-interacting agegraphic dark energy model, the reasonable
behavior of $\Omega_D$ can be obtained by solving the evolving
equations numerically with $n=3$ and the initial condition
$\Omega_{D0}=0.73$.

\section*{Acknowledgments}
This work is supported by Research Fund for the Doctoral Program of
Higher Education of China under Grant No. 20106101120023.

\end{document}